\documentclass[3p,times,procedia]{elsarticle}
\usepackage{nupha_ecrc}

\volume{00}

\firstpage{1}

\journalname{Nuclear Physics A}

\runauth{C. Bierlich}

\jid{nupha}

\jnltitlelogo{Nuclear Physics A}

\usepackage{amssymb}
\usepackage[utf8]{inputenc}

\usepackage[figuresright]{rotating}

\newcommand{\pythia}{P\protect\scalebox{0.8}{YTHIA}8}

\newcommand{\dipsy}{\protect\scalebox{0.8}{DIPSY}}

\begin{document}

\begin{frontmatter}



\dochead{XXVIIth International Conference on Ultrarelativistic Nucleus-Nucleus Collisions\\ (Quark Matter 2018)}

\title{Microscopic collectivity: The ridge and strangeness enhancement from string--string interactions}
\author[lund,bohr]{Christian Bierlich\corref{author}}

\cortext[author] {\textit{E-mail address:} christian.bierlich@thep.lu.se. Preprint number: LU-TP 18-22, MCnet-18-16. Work done in collaboration with G. Gustafson and L. L\"onnblad. Work funded in part by the Swedish Research Council, contracts number 2012-02283 and 2017-0034, and in part by the MCnetITN3 H2020 Marie Curie Initial Training Network, contract 722104. Travel support from Bokelunds resestipendiefond is gratefully acknowledged.}
\address[lund]{Dept. of Astronomy and Theoretical Physics, S{\" o}lvegatan 14A, S-223 62 Lund, Sweden}
\address[bohr]{Niels Bohr Institute, Blegdamsvej 17, 2100 Copenhagen, Denmark}

\begin{abstract}
We present the microscopic model for collective effects, built on string--string interactions (string shoving and rope hadronization), recently implemented in the \pythia~and \dipsy~event generators. Rope hadronization is shown to give a good description of strangeness enhancement across pp, pA and AA collision systems, while string shoving qualitatively describes the ridge observed in pp collisions.
\end{abstract}


\end{frontmatter}


\section{Introduction}
\label{sec:intro}
The basis of hadronization in the \pythia~event generator \cite{Sjostrand:2014zea} is the Lund string model \cite{Andersson:1983jt,Andersson:1979ij}, where the confined colour field between a $q\bar{q}$ pair is modelled as a semi-classical string with tension $\kappa = 1$ GeV/fm. The string breaks into smaller pieces in new $q\bar{q}$ breaking, when energetically favorable. Longitudinal momenta are determined by the Lund fragmentation function:
\begin{equation}
	\label{eq:lundff}
	f(z) \propto z^{-1}(1-z)^a \exp\left(\frac{-bm_\perp}{z} \right),
\end{equation}
where $z$ is the \textit{remaining} available momentum at the $i$'th breaking, and $a$ and $b$ are tunable parameters. The transverse momentum is determined by:
\begin{equation}
	\label{eq:sch}
	\frac{d\mathcal{P}}{dp_\perp} \propto \kappa \exp\left( -\frac{\pi m^2_\perp}{\kappa} \right).
\end{equation}
The mass is the mass of the quarks or di--quarks produced pairwise in the breaking. If one inserts constituent quark masses $m_u \approx 0.3$ GeV, $m_s \approx 0.5$ GeV to calculate strange quark surpression directly, one obtains a suppression factor of 0.08, which is too low to describe data. In practise the strange quark and di-quark supression factors are therefore treated as free parameters, fitted to LEP data.

When extended to $q\bar{q}g$ topologies (in $e^+e^-$), one string piece is stretched between the quark and the gluon, and another between the gluon and the anti-quark, giving rise to the ``string effect" observed by JADE \cite{Bartel:1981kh}, a long range (in rapidity) effect. 

Since the Lund string gives rise to non--trivial correlations in rapidity already in $e^+e^-$ collisions, it is relevant to ask to what degree such effects are also present in pp collisions. Here the partonic final state is generated through a more complicated multiparton intereractions (MPI) scenario \cite{Sjostrand:1987su}. The mapping from partonic final state to Lund strings it not unique in this case, but Colour Reconnection (CR) models reshuffle the strings in order to minimize the free energy. It has been noted that CR gives rise to short range collective effects \cite{Ortiz:2013yxa}.

\section{The microscopic model for collectivity}
If long range collective phenomena, such as the pp ridge \cite{Khachatryan:2010gv}, are to be included in MPI + shower + string picture outlined above, further steps need to be taken. In a series of papers \cite{Bierlich:2014xba,Bierlich:2015rha,Bierlich:2016vgw,Bierlich:2017vhg}, we have proposed a microscopic model for collectivity, based on interacting Lund strings. The model relies on temporal evolution of an event, outlined as ($\tau$ is the string eigentime):\\
$\tau \approx 0$ fm/c: Strings have no transverse extension. No interactions, partons may propagate.\\
$\tau \approx 0.6$ fm/c: Parton shower ends. Depending on diluteness, strings may shove each other around.\\
$\tau \approx 1$ fm/c: Strings at full transverse extension. Shoving effect maximal.\\
$\tau \approx 2$ fm/c: Strings will hadronize. Possibly as a colour rope.\\

\begin{figure}
	\begin{center}
	\includegraphics[width=0.43\textwidth]{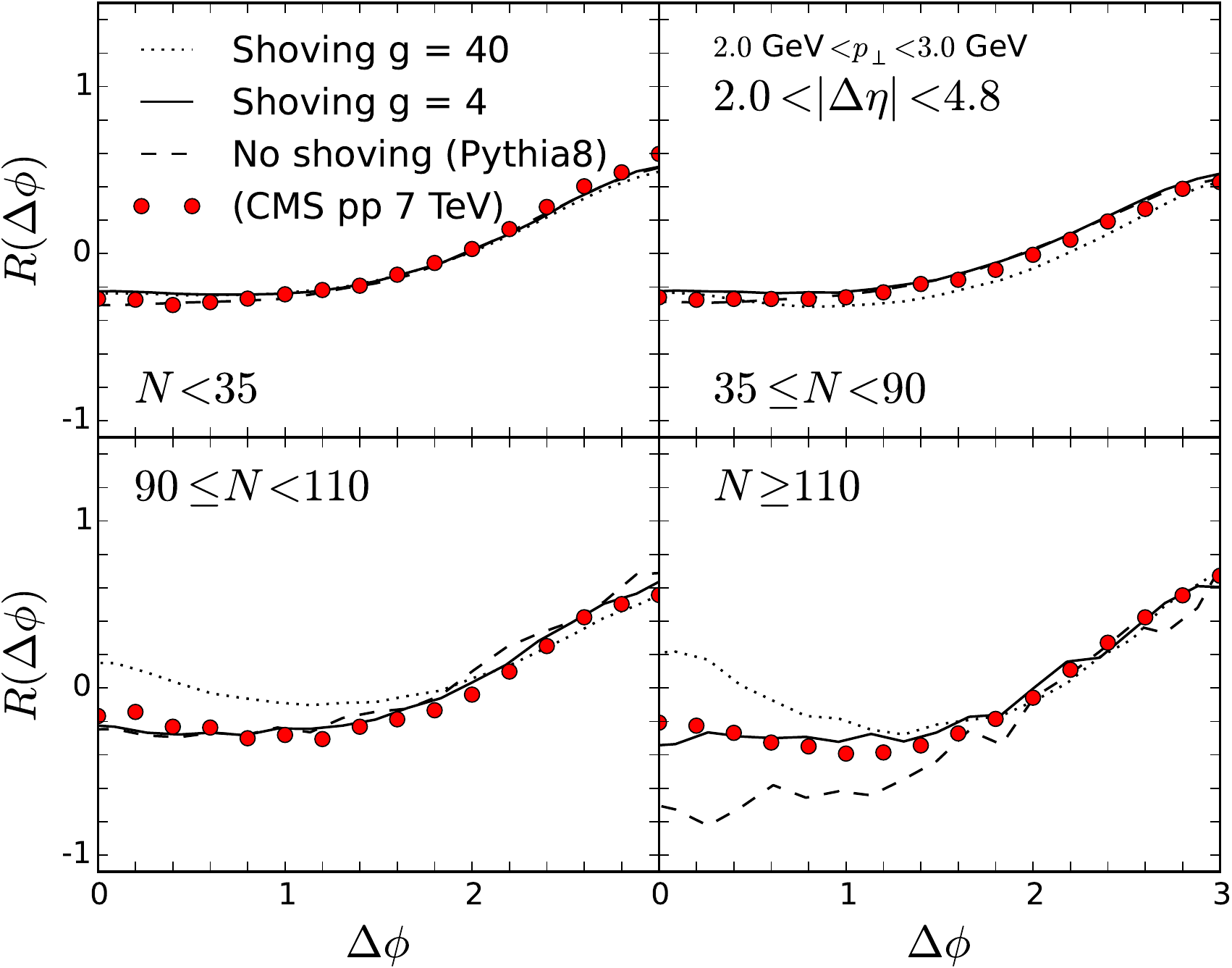}
	\includegraphics[width=0.43\textwidth]{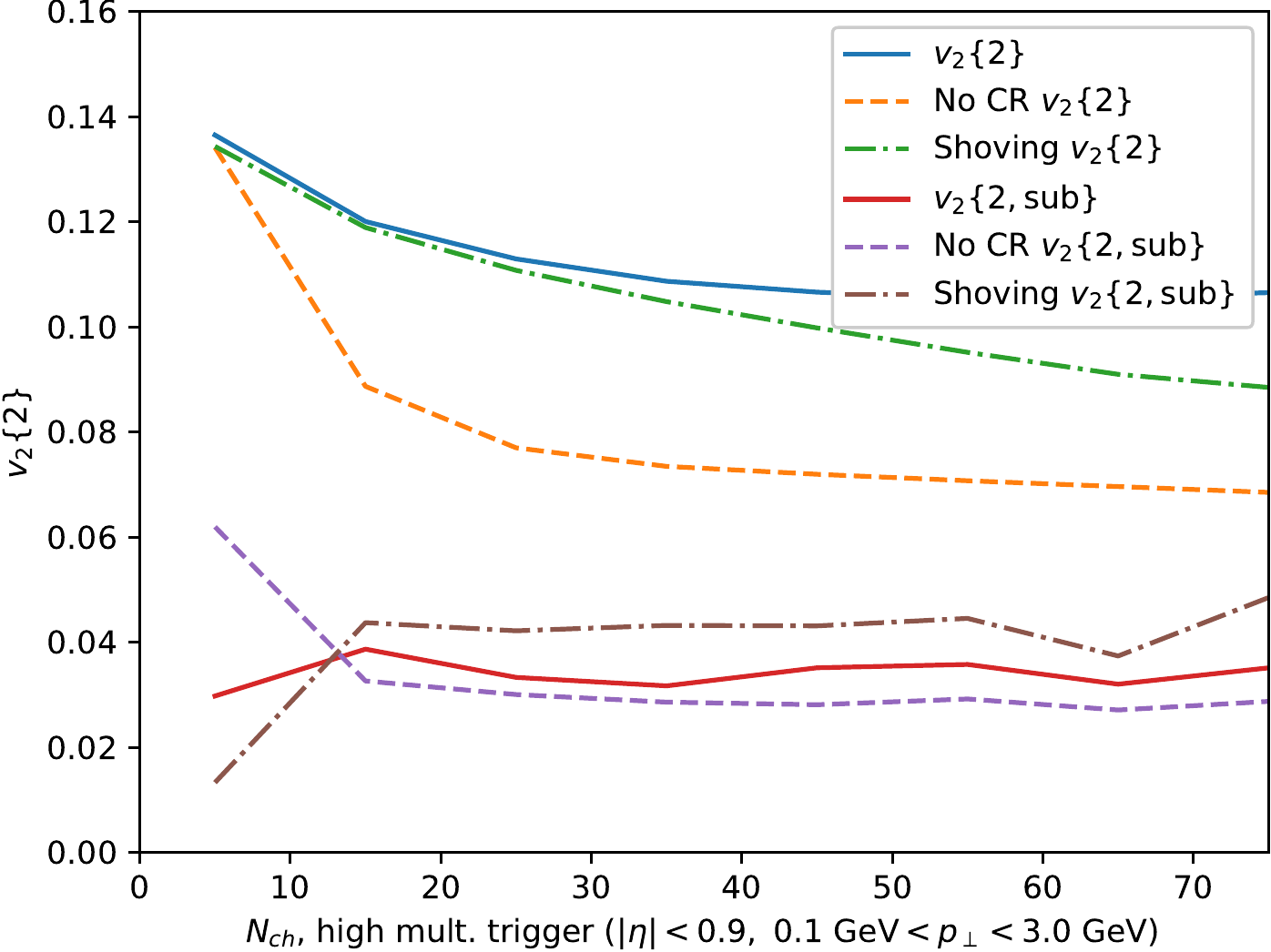}
	\end{center}
	\caption{\label{fig:shoving} (left) The appearance of a ``ridge" in pp at high multiplicities, when enabling shoving in \pythia. (right) Flow--coefficients $v_2\{2\}$ with and without a $\Delta \eta$--gap without CR, with CR and with shoving.}
\end{figure}

\subsection{String shoving}
When the energy density at times $\tau \approx 0.6-2$ fm/c is high enough, the overlapping strings generate a transverse pressure, which manifests itself as a transverse flow \cite{Abramovsky:1988zh}. The (classical) force between two string pieces is calculated by treating the string in analogy with a colour superconductor, with a field strength close to a Gaussian \cite{Cea:2014uja}: $\mathcal{E}(r_\perp)=C\,\exp(-r_\perp^2/2R^2)$, where $r_\perp$ is the distance from the string core, and $R$ specifies its width. This results in a force per unit length between two strings at distance $d_\perp$ apart, given by:
\begin{equation}
 f(d_\perp) = \frac{g\kappa
  d_\perp}{R^2}\exp\left(-\frac{d^2_\perp(t)}{4R^2} \right).
\label{eq:repulsion}
\end{equation}

The model is implemented in \pythia, given a simplistic model for MPI distributions in impact parameter space (see ref.~\cite{Bierlich:2017vhg} for details). In fig. \ref{fig:shoving} (left) it is shown how the ridge is seen to emerge at high pp multiplicities, with a strength parameter $g$ (\textit{cf.} eq. (\ref{eq:repulsion})) not too far from unity. In fig. \ref{fig:shoving} (right) $v_2\{2\}$ flow coefficients with and without a $\Delta \eta$--gap \cite{Bilandzic:2013kga} are shown without CR, with CR (\pythia~default) and with shoving. It is seen that CR gives rise to a large $v_2\{2\}$, which is removed by adding a gap. Some contribution from shoving is, however, seen to survive.
Results are expected to improve, if interfaced with a more realistic model for MPI distributions, \textit{eg.} proton hotspots \cite{Albacete:2016pmp}.

\subsection{Rope hadronization}
At $\tau \approx 2$ fm/c, strings hadronize. If the density is still high enough, strings may hadronize as colour multiplets with a higher effective string tension, called colour ropes. This idea was originally proposed by Biro \textit{et al.} \cite{Biro:1984cf}, and pursued by many groups (see ref.~\cite{Bierlich:2014xba} and references therein). It is seen directly from eq. (\ref{eq:sch}), that increasing the string tension will increase the likelihood of producing heavier quarks such as $s$. The $p_\perp$ and $m_q$ parts factorize, and the strangeness suppression is given in terms of the quark masses as:
\begin{equation}
	\label{eq:sup}
	\rho = \frac{\mathcal{P}_s}{\mathcal{P}_{u,d}} = \exp\left(-\frac{\pi(m^2_s - m^2_u)}{\kappa}\right).
\end{equation}

\begin{figure}
	\begin{center}
		\includegraphics[width=0.47\textwidth,clip]{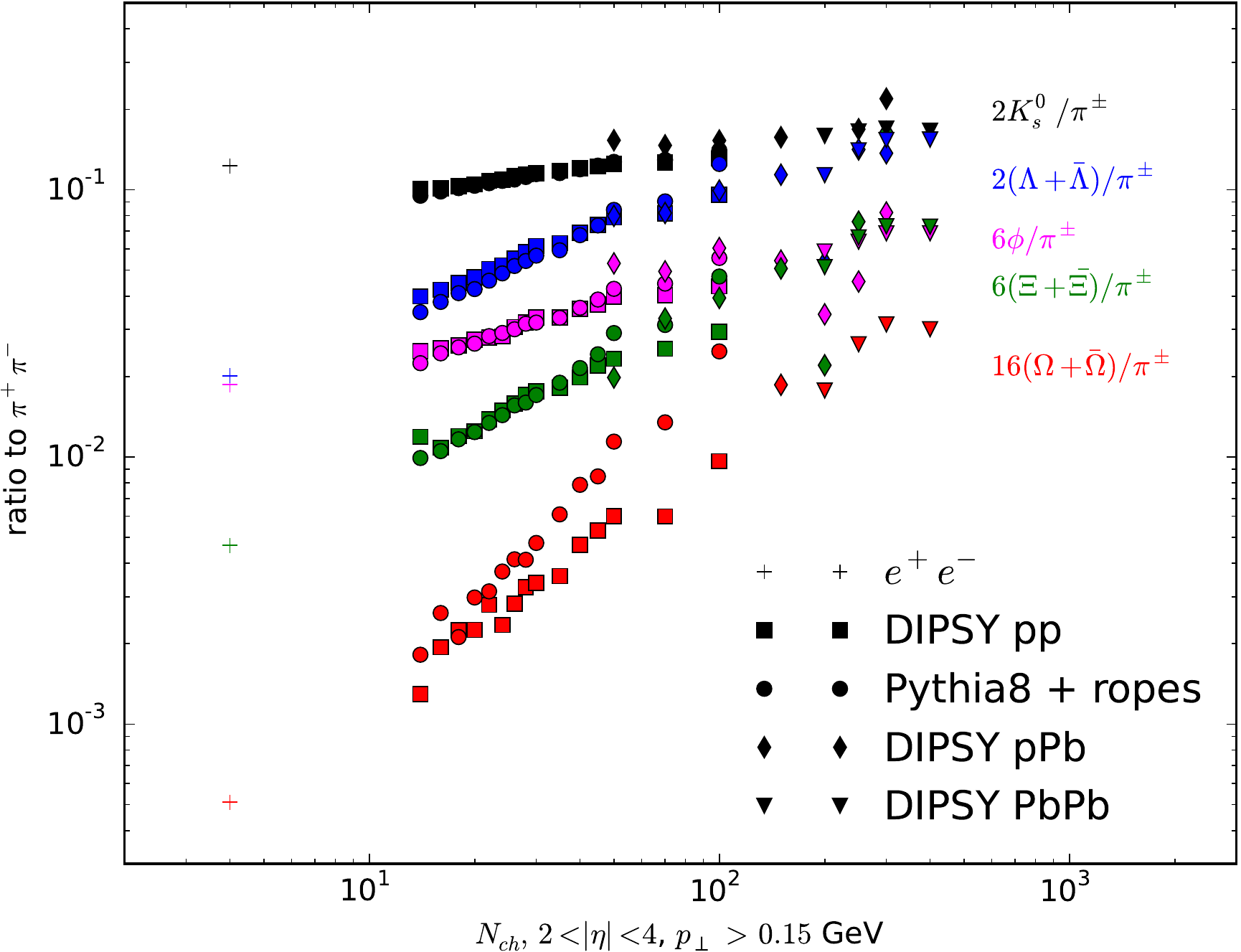}
		\includegraphics[width=0.43\textwidth]{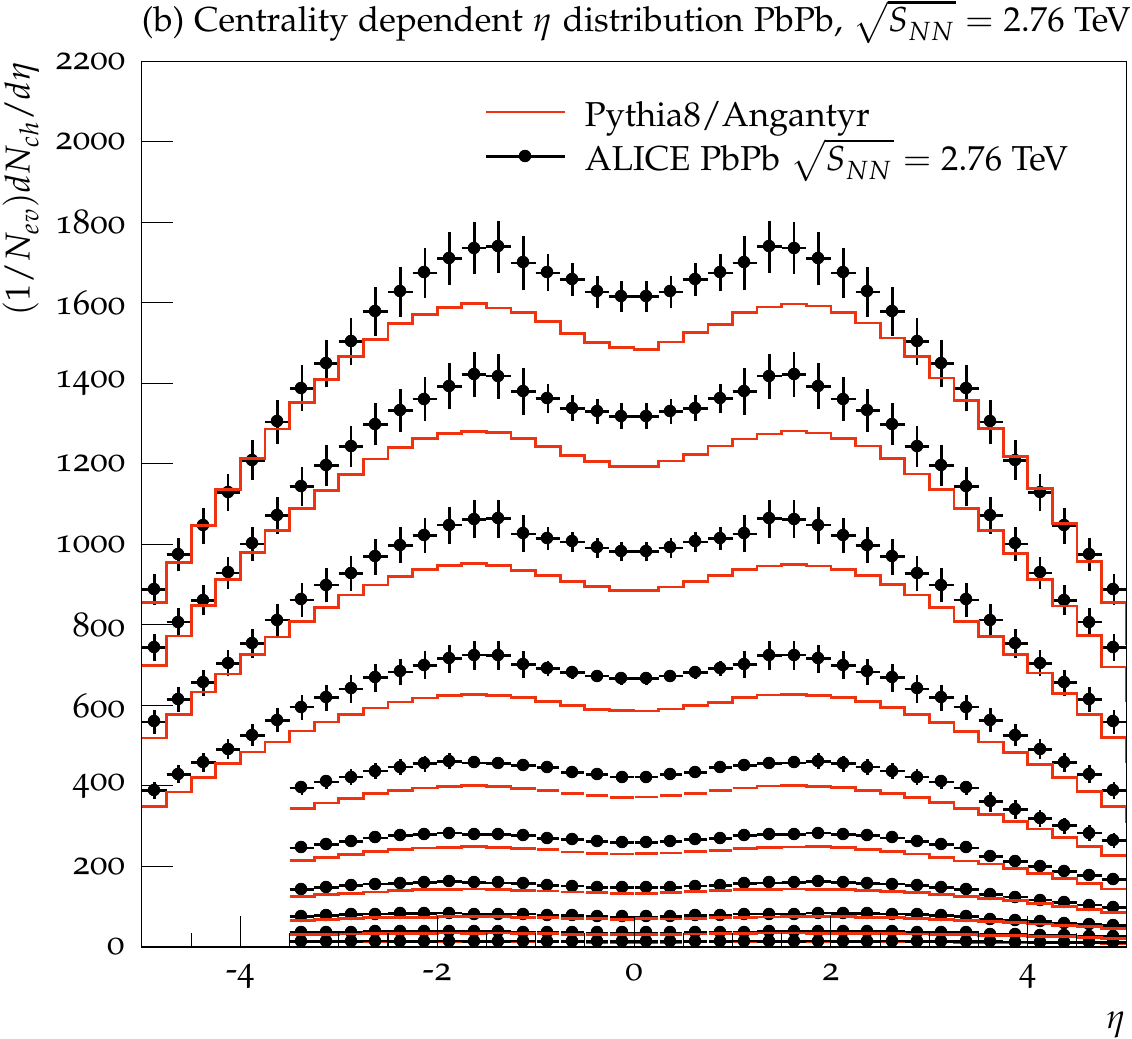}
	\end{center}
	\caption{\label{fig:ropefig}(left) Ratios of strange particles to pions across $e^+ e^-$, pp, pPb and PbPb, as function of forward charged multiplicity with Rope Hadronization as implemented in \pythia and \dipsy. (right) Charged particle density as function of $\eta$ in centrality bins, as calculated using the Angantyr extension to \pythia.}
\end{figure}

In figure \ref{fig:ropefig} (left) the increase in the ratio of strange/non-strange particles as function of multiplicity (centrality) with rope hadronization is shown. This was measured by ALICE \cite{ALICE:2017jyt}. In that paper, comparisons with the \dipsy~Monte Carlo event generator \cite{Flensburg:2011kk}, including the same rope model as introduced above, was shown. In the ALICE analysis, good agreement between data and \dipsy~for $K^0_s/\pi$ and $\Lambda/\pi$ was shown, while agreement with $\Xi/\pi$ and $\Omega/\pi$ is somewhat worse. Comparing the ratios in pp produced by \pythia~and \dipsy~respectively (in figure \ref{fig:ropefig} (left)), it is seen that the $\Xi/\pi$ and $\Omega/\pi$ ratios are improved in the \pythia~implementation. The $p/\pi$ ratio (not shown) is, as remarked also in ref.~\cite{ALICE:2017jyt}, not well described by the rope model. In the same figure the ratios for pPb and PbPb are also shown, using \dipsy, showing a similar continuous transition among colliding systems as data. The microscopic model for collectivity as implemented in \pythia~is currently being further developed to describe also pPb and PbPb final state, using the Angantyr \cite{Bierlich:2016smv,Bierlich:2018xfw} model for heavy ion collisions. The latest developments in the Angantyr model are shown in figure \ref{fig:ropefig} (right). Angantyr uses a Glauber initial state calculation, including additional fluctuations, and maps proton--proton collisions onto that. Without including any additional nuclear effects, final state multiplicity in PbPb collisions is reproduced well, as shown. As an impact--parameter picture is included from the Glauber initial state, the Angantyr model is suitable for introducing ropes and shoving formulated in impact--parameter space.

\section{Outlook}
Event generators like \pythia~are often used in heavy ion physics as an estimate of the ``no QGP" case, especially in small systems. The phenomenological models implemented have, however, only recently started being extended towards handling cases where many strings interact in a single pp event. Such effects ought to be taken into account when using general purpose Monte Carlo event generators for special purpose cases as these. We have shown recent developments of string shoving and rope hadronization in the framework of a microscopic model for collectivity. These are not the only possible venues. More formal developments in the effects of QCD coherence on collectivity may provide valuable guidelines \cite{Blok:2017pui}. Furthermore development in CR models \cite{Christiansen:2015yqa} are promising in terms of at least baryon enhancement, as are similar developments in the cluster hadronization model \cite{Gieseke:2017clv}.





\bibliographystyle{elsarticle-num}
\bibliography{qm.bib}







\end{document}